\documentclass[apj]{emulateapj}
\begin{document}
\title{General Relativistic Binary Merger Simulations and Short Gamma Ray Bursts}
\author{Joshua A. Faber
\altaffilmark{1}
\altaffilmark{2}
\email{jfaber@uiuc.edu}}
\author{Thomas W. Baumgarte
\altaffilmark{3}}
\author{Stuart~L.~Shapiro 
\altaffilmark{1}
\altaffilmark{4}}
\author{Keisuke Taniguchi
\altaffilmark{1}}
\altaffiltext{1}{Department of Physics, University of Illinois at
  Urbana-Champaign, Urbana, IL 61801}
\altaffiltext{2}{NSF Astronomy and Astrophysics Postdoctoral Fellow}
\altaffiltext{3}{Department of Physics and Astronomy, Bowdoin College,
  Brunswick, ME 04011}
\altaffiltext{4}{Also Department of Astronomy and NCSA, UIUC, Urbana,
  IL 61801}

\begin{abstract}
The recent localization of some short-hard gamma ray bursts (GRBs) in
galaxies with low star formation rates has lent support to the
suggestion that these events result from compact object binary
mergers.  We discuss how new simulations in general relativity 
are helping to identify the central engine of short-hard GRBs.  Motivated by 
our latest relativistic black hole-neutron star merger calculations, we discuss 
a scenario in which these events may trigger short-hard GRBs, 
and compare this model to competing relativistic models involving binary 
neutron star mergers and the delayed collapse of hypermassive neutron 
stars.  Distinguishing features of these models may help guide future 
GRB and gravitational wave observations to identify the nature of the 
sources.
\end{abstract}
\maketitle

\section{Introduction and Observations}

Gamma-ray bursts are short duration phenomena characterized by
time-varying, high-energy, non-thermal electromagnetic emission.
Based on their duration and energy spectra they are typically
categorized into two categories: ``short/hard'' and ``long/soft''
\citep{Kouveliotou}.  Many optical and X-ray counterparts of long
bursts have been seen, some in coincidence with Type Ib/c supernovae,
representing extremely energetic collapses of massive stars
\citep{Paczynski98,Hjorth03}.  By contrast, it has been much more
difficult to identify counterparts for short GRBs (SGRBs), except for
the known soft gamma repeaters, which may be observed as SGRBs
\citep{Palmer}, but cannot explain more than a small fraction of the
observed SGRB sample \citep{LGG,NGPF}.

Recently, the {\it Swift} and {\it HETE-2} satellites have localized
for the first time the X-ray afterglow following short-period GRBs:
GRBs 050509b \citep{Gehrels}, 050724 \citep{Barth,Berger},
050813 \citep{Berger2}, and 051221 \citep{Soder} by {\it
Swift}; GRB 050709 \citep{Hjorth,Fox} by {\it HETE-2}.  Details about
the physical parameters observed and inferred from these bursts can be
found in \citet{Berger3}.  In all cases, the inferred host had a
rather low star-formation rate.  This finding disfavors the collapse of a
massive star as a progenitor, since those systems have very short
lifetimes.  Instead, it favors the identification of a compact object
binary merger as the progenitor, as originally suggested by
  \citet{Pac86}; a significant fraction of both neutron star-neutron
star (NSNS) and black hole-neutron star (BHNS) binaries will take longer
than 1 Gyr between
formation and merger \citep{BBR}.  Other scenarios, such as the
accretion-induced collapse of a NS with a WD companion have been
proposed \citep{Dermer}, but have not been studied numerically in as
much detail.

The observed characteristics of SGRBs can be examined in light of
population synthesis calculations for compact object binaries, but it
is difficult to disentangle BHNS vs. NSNS merger scenarios given the
current observational and theoretical uncertainties.  The inferred
  rate of observed SGRBs and the merger 
rate of their putative sources are consistent, but the latter are uncertain by
at least two orders of magnitude: ${\cal R}_{GRB}\equiv
(\Omega/4\pi){\cal R}_{m}\equiv (\theta^2/2){\cal R}_{m}$, where
$\theta$ is the beaming angle.  The most recent GRB rate estimates
yield ${\cal R}_{GRB}\sim 1-3~{\rm Myr^{-1}}$ per Milky Way Galaxy
\citep{Guetta2,Nakar}, over an order of magnitude larger than previous
results \citep{Guetta, Ando}. Breaks in the observed x-ray spectra
imply $\theta\sim 0.2-0.3$, and thus $\Omega/4\pi=0.02-0.05$
\citep{Soder}. The predicted compact object merger rates per
  Milky Way Galaxy fall in the
range ${\cal R}_{m}\sim 1-100~{\rm Myr^{-1}}$ for
NSNS and ${\cal R}_{m}\sim 0.5-10~{\rm Myr^{-1}}$ for BHNS
\citep{VT,BBR,KKL}.  The rates are uncertain by at least an order of
magnitude and the beaming angle by a factor of a few.  An extremely
high SGRB rate may indicate that at least some SGRBs are the products
of NSNS mergers, as these are much more common than BHNS mergers in
population synthesis calculations \citep{VT,BBR}.

The projected distance from a localized SGRB to the center of its host
galaxy is insensitive to whether the merging binary is an NSNS or BHNS
system.  Recent results show the respective projected distance
likelihood functions are nearly the same for large galaxies, and have
only minor differences in smaller ones, which would rule out positive
identifications based on location alone \citep{Bel06}.

The range of fluxes seen from short GRBs is rather large, indicating that
the burst energy must depend sensitively on at least one
parameter of the progenitor system.  The observed fluxes of short GRBs
typically fall in the range $5\times 10^{-6}-10^{-4}~{\rm
erg~cm}^{-2}~{\rm s}^{-1}$, and fluences $10^{-7.5}-10^{-5.5}~{\rm
erg~cm}^{-2}$ \citep{Balazs}.  Assuming a characteristic distance of
$1~{\rm Gpc}$, we see that the luminosities $L$ and total energy
released $E$ for the burst satisfy
$(4\pi/\Omega)L=10^{50.5}-10^{52}~{\rm erg~s^{-1}}$ and
$(4\pi/\Omega)E=10^{48.5}-10^{50.5}~{\rm erg}$, respectively.

The first localized short burst, GRB 050509b, has an extremely low
measured isotropic energy in $\gamma$-rays, $E_\gamma=3\times
10^{48}(\Omega/4\pi)~{\rm erg}$, compared to previously observed short
GRBs.  Assuming that the GRB is at the measured redshift of $z=0.225$,
the luminosity of the burst is actually similar to the other localized
bursts, $\sim 10^{50}~{\rm erg~s^{-1}}$, but it lasted for a much
shorter time.  In general, as we see from Table 1 of \citet{JMAP},
there is significantly smaller variation in the peak luminosity than
in the isotropic energy output.

We note that the low energy for GRB 050509b argues for an extremely
low density of baryons surrounding the GRB.  Typically, it is assumed
that the Lorentz factor of the jet $\Gamma$ will be no bigger than the
ratio $\eta$ of the energy in the jet to the rest energy of the
baryons through which it must travel, i.e. $\Gamma \le \eta\equiv
E_{\gamma}/M_{b}c^2$ \citep{ShemiPiran}.  For GRB 050509b, the least
energetic of the observed bursts, assuming the Lorentz factor
$\Gamma\gtrsim 100$ \citep{Oech} implies that at most $10^{-8}
(\Omega/4\pi) M_{\odot}$ of material can surround the progenitor.
These results are confirmed by the numerical calculations of
\citet{Aloy}, who find that the density of baryons surrounding the
burst must fall rapidly with increasing radius.  For sufficiently
large baryonic loading, they find the observational counterpart to a
merger is not a GRB.

\section{Progenitor models: theoretical and observational constraints}

Standard models for merger-induced short-duration GRBs involve
hot, massive accretion disks ($M>0.01 M_{\odot}$) around spinning
BHs (see \citealt{Piran} for a thorough review).  As the disk
is accreted, one of two possible mechanisms is responsible for the
creation of a gamma-ray jet.  One suggestion is that the hot material
emits neutrino-antineutrino pairs that annihilate above and below the
disk to produce a relativistic jet containing $e^{-}-e^{+}$ pairs and
photons (see, e.g. \citealp{Pac86,GDN}).  Calculations of accreting
disks have shown that they can generate sufficient energy to power a
gamma-ray burst \citep{PWF}.  The viscosity in the disk primarily
determines the timescale of the resulting burst \citep{NPK}, while the
mass of the BH may determine the overall energy scale \citep{PWF}.
These results are supported by \citet{SRJ1,SRJ2}, who find that the
viscosity (and to a lesser extent, BH spin and disk mass) are
responsible for determining the overall burst energy. Roughly
speaking, luminosities of up to $10^{50}~{\rm erg~s}^{-1}$ can be
produced, so long as the disk is sufficiently massive (representing
$~5\%$ of the total binary mass) and the effective viscosity
sufficiently strong ($\alpha=0.1$).

Alternatively, general
relativistic (GR) magnetohydrodynamic (MHD) effects may allow
infalling matter to tap the spin energy of a BH via the
Blandford-Znajek effect to produce an energetic jet \citep{BZ}.  In
this case, the crucial parameters to determine the energetics are the
mass accretion rate and the spin of the BH.  A review of GRB models
tapping spin energy can be
found in \citet{ZM}.

\subsection{Neutron Star-Neutron Star Binaries and Hypermassive Remnants}

There has been a great deal of numerical work performed to study
merging NSNS binaries, including recent calculations performed using
fully GR hydrodynamics and physically realistic nuclear equations of
state (EOS; \citealp{STU,ST}).  From these results, two separate channels have
been identified which could lead from a merger to the production of an
SGRB.

If the total binary mass is below some critical threshold, $M_{thr}$,
a differentially rotating hypermassive neutron star (HMNS) can
be formed that is initially stable against gravitational collapse,
even though its mass is above the maximum value for uniformly
rotating configurations \citep{BSS,Morrison}.  Delayed gravitational
collapse follows, triggered either by gravitational wave
dissipation if the remnant forms a bar, or by magnetic redistribution
of angular momentum \citep{BSS,Shap00,CSS,Duez,Duez06,ShiUIUC}.  
The numerical simulations of Shibata et al.~use a stiff, realistic EOS 
\citep{APR}, for which the maximum mass is slightly higher than the
highest measured pulsar mass, $M=2.1\pm0.2~{\rm
M_{\odot}}$ for PSR J0751-1807 \citep{Nice}.  These simulations 
set a lower limit for the
critical mass $M_{thr}\gtrsim 2.7 M_{\sun}$.  This 
exceeds all known NSNS binaries containing a radio pulsar, with the
possible exception of PSR 1913+16 ($M_{\rm tot} = 2.83~{\rm
M_{\odot}}$, \citealp{Stairs}).  Thus, it appears that HMNS formation is
likely in most merging NSNS binaries.  MHD simulations in full
GR show that the HMNS undergoes a delayed collapse, resulting in a
hot, magnetized torus surrounding a rotating BH, together with a
magnetic field collimated along the polar axis.  These conditions are
  favorable for a burst powered either by neutrino annihilation or MHD
  effects \citep{Duez,Duez06,ShiUIUC}.

Alternatively, for higher-mass
NSs, mergers with binary mass ratios sufficiently far from unity might
lead to the formation of a relatively massive disk around a BH formed
promptly during the merger \citep{ST}.  For mass ratios $q\sim 0.7$,
disk masses of $0.01-0.1 M_{\odot}$ are possible, whereas more
equal-mass mergers produce much lower mass disks, with insufficient
thermal energy to power an SGRB \citep{STU}.

The key unanswered question for the HMNS scenario is whether ``baryon
pollution'' from merger ejecta along the polar axis can be cleared out
of the funnel through which the presumed GRB jet will propagate.
Relativistic numerical calculations \citep{Duez,Duez06,ShiUIUC}
indicate that MHD effects seem to be sufficient to accomplish this
task, producing baryon densities along the polar axis that satisfy the
constraints described in \citet{ShemiPiran}.  For the unequal-mass
NSNS merger case, the polar axis is essentially free of intervening
matter, but the likelihood of this scenario depends on how far
from unity binary mass ratios are likely to fall, as all observed
systems containing radio pulsars have $q\gtrsim 0.9$.  Systems with
components too close together in mass produce significantly lower
mass disks, which lack the required neutrino luminosity to power
the SGRB \citep{ST}.

\subsection{Black Hole-Neutron Star Binaries}\label{sec:bhns}

The defining parameter for determining the qualitative evolution of a
BHNS merger is the binary mass ratio $q\equiv M_{\rm NS}/M_{\rm BH}$.
Assuming that the tidal disruption of the NS begins at a separation
$a_{\rm R}$ where its volume in
isolation is equal to the volume of its Roche lobe,
\begin{equation}
a_{\rm R} / M_{\rm BH} = 2.17q^{2/3}(1+q)^{1/3}{\cal C}^{-1} \label 
{eq:ar}
\end{equation}
for a NS of compactness ${\cal C}\equiv M_{\rm NS}/R_{\rm NS}$ \citep 
{PacRoche},
using geometrized units with $G=c=1$.
For a sufficiently large BH (and thus small $q$) $a_{\rm R}$
is smaller than the innermost stable circular orbit (ISCO) $a_{\rm
ISCO}$, so that the NS passes through the ISCO before being disrupted
tidally.   For a typical NS of compactness ${\cal C}=0.15$ the  
critical mass
ratio at which tidal disruption occurs at $a_{\rm ISCO}$ is  
approximately $q=0.24$.
It has been argued that even for this "fiducial" binary the
rapid plunge timescale at the ISCO might prevent significant
disk formation \citep{Miller}.
Our simulations suggest, however, that rapid angular momentum  
transfer during
tidal disruption ejects a significant fraction of the matter back  
outside
the ISCO, leading to a sizable disk.  Clearly, a detailed understanding
of this dynamical process requires an accurate description of the  
strong field BHNS spacetime, as we discuss below.

We performed 3+1-dimensional smoothed particle hydrodynamics (SPH)
calculations of BHNS mergers in the conformal flatness (CF) approximation to
GR \citep[hereafter FBSTR]{BH1}.  For a Schwarzschild BH, our
scheme identifies including the relativistic ISCO exactly
and accounts for relativistic dynamics within the ISCO, unlike
previous pseudo-Newtonian calculations \citep{LK,JERF,RSW,Ross}.  We
adopt the relativistic, irrotational binary models of \citet{TBFS} as
initial data.  Our 
adiabatic evolution scheme is the same as that described in FBSTR, but
we solve the coupled non-linear elliptic field equations of the CF
scheme using an iterative FFT-convolution based solver and add
artificial viscosity to the hydrodynamical equations in order to study
the shock heating of the material during the merger process (compare
\citealt{HKAV}).  Both the initial data
and the evolution scheme represent the state-of-the-art for including
NS self-gravity self-consistently within a GR
hydrodynamical formalism.

We consider two models, both containing $n=1$ polytropic NS.  First is
a NS of compactness $0.15$, denoted ``case A''.  As in FBSTR, our
current method is limited to extreme mass ratios, so we choose
$q=0.1$, for which $a_{\rm R}=3.2 M_{\rm BH}$ according to
Eq.~\ref{eq:ar}.  We cannot directly simulate the fiducial case 
($q=0.24$, ${\cal C}=0.15$) under these
assumptions, so we take an alternate approach for our second
calculation.  Simultaneously reducing the mass ratio to $q=0.1$ and
the NS compactness to ${\cal C}=0.09$ leaves $a_{\rm R}/M_{\rm BH}$
virtually unchanged.  From the argument above, this binary, denoted
``case B'' should tidally disrupt slightly within the ISCO, mimicking
the dynamics of the fiducial binary.

\begin{figure*}
\plotone{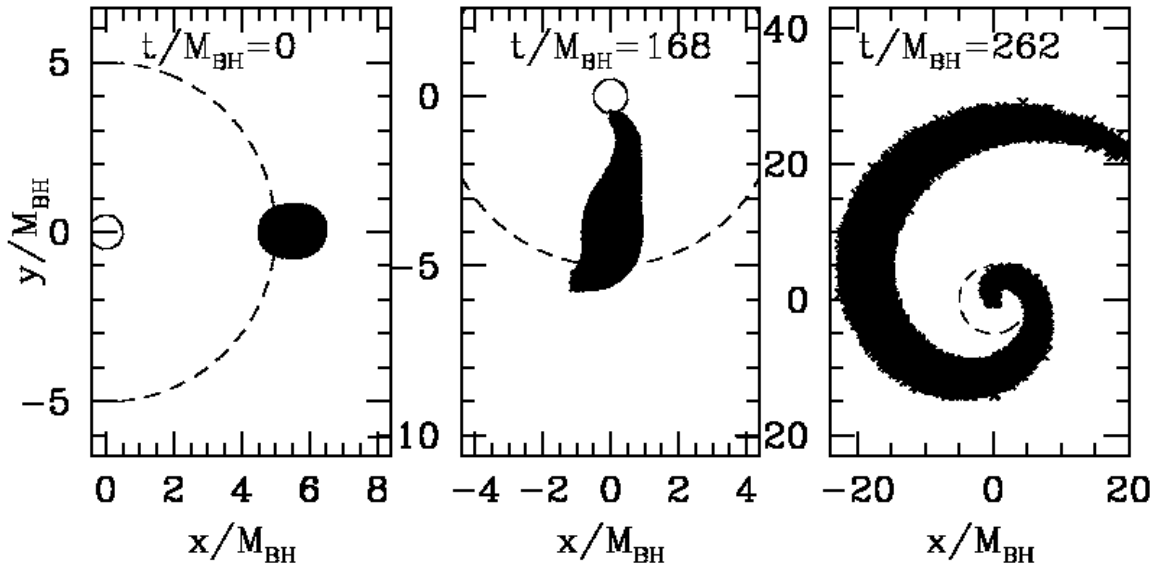}
\caption{Initial (left), early (center) and intermediate (right)
  configurations of the NS in case B, projected into the orbital
  plane. The NS has a compactness $M/R=0.09$ and the binary mass ratio
  is $M_{\rm NS}/M_{\rm BH}=0.1$.  We see the NS disrupts near the
  ISCO (dashed curve) to produce a single mass-transfer stream, which
  eventually wraps around the BH (solid curve) to form a torus.  The
  initial orbital period is $P=105M_{\rm BH}$. \label{fig:xyz}}
\end{figure*}

The evolution of case A is straightforward and probably uninteresting
as an SGRB source: the entire NS spirals toward the BH, and no matter
is ejected outside the ISCO to form an accretion disk.  Case B, on the
other hand, does lead to the formation of an accretion disk, as we
show in Fig.~\ref{fig:xyz}.  Note that radii are measured in isotropic
coordinates, in which the BH radius is $r_{\rm BH}=0.5M_{\rm BH}$ and
the ISCO radius $a_{ISCO}\approx 5M_{\rm BH}$.  We see that while the
NS inspirals until $98\%$ of its mass lies within the ISCO, rapid
redistribution of angular momentum during tidal disruption causes an
outwardly directed spiral arm to form, sending some matter back
outside the ISCO.  

Eventually, the BH accretes $M_{acc}\approx 0.75 M_{\rm NS}$ directly,
while part of the remaining mass forms a disk of mass $M_{disk}\approx
0.12M_{\rm NS}$ and part is ejected completely from the system
($M_{ej}\approx 0.13M_{\rm NS}$).  Here we distinguish bound and
unbound trajectories by the sign of $u_0-1$, where $u_0$ is the
time-component of the matter 4-velocity, which remains nearly constant
in time for outflowing gas on approximately geodesic trajectories.  We
note that this configuration satisfies all the geometric constraints
required for a GRB progenitor, as all matter lies in the equatorial
plane rather than the polar axis.

\begin{figure*}
\plotone{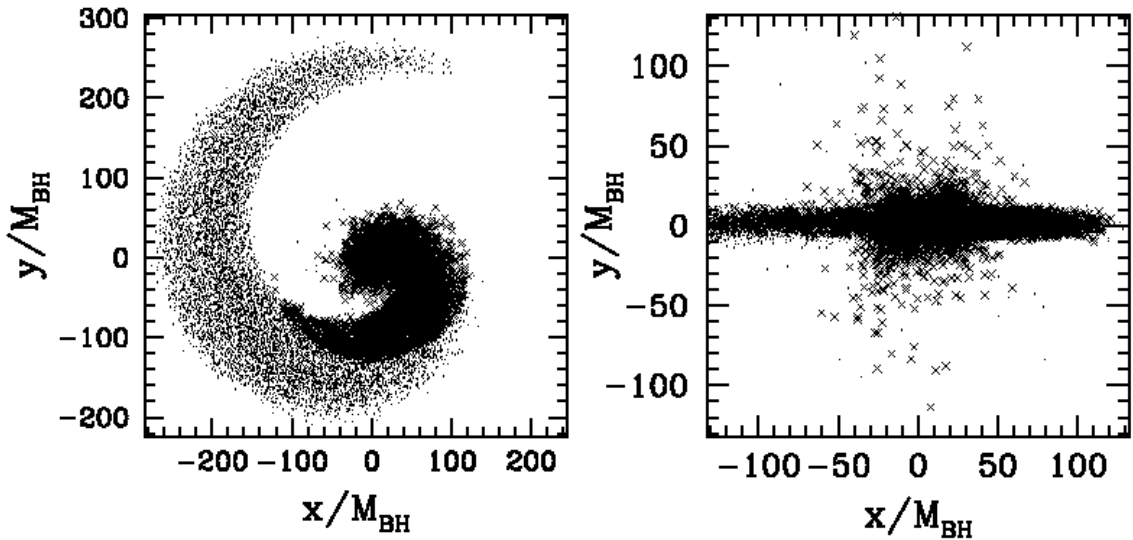}
\caption{Matter configuration at the end of the simulation,
  $T=990M_{\rm BH}$, projected onto the equatorial (left panel) and
  meridional (right panel) showing the hot torus located within
  $r<50M_{\rm BH}$.  Bound fluid elements (satisfying $u_0-1<0$) are
  shown as crosses, unbound elements as points.  Note the different
  scales.\label{fig:xyz2}}
\end{figure*}

\begin{figure*}
\plotone{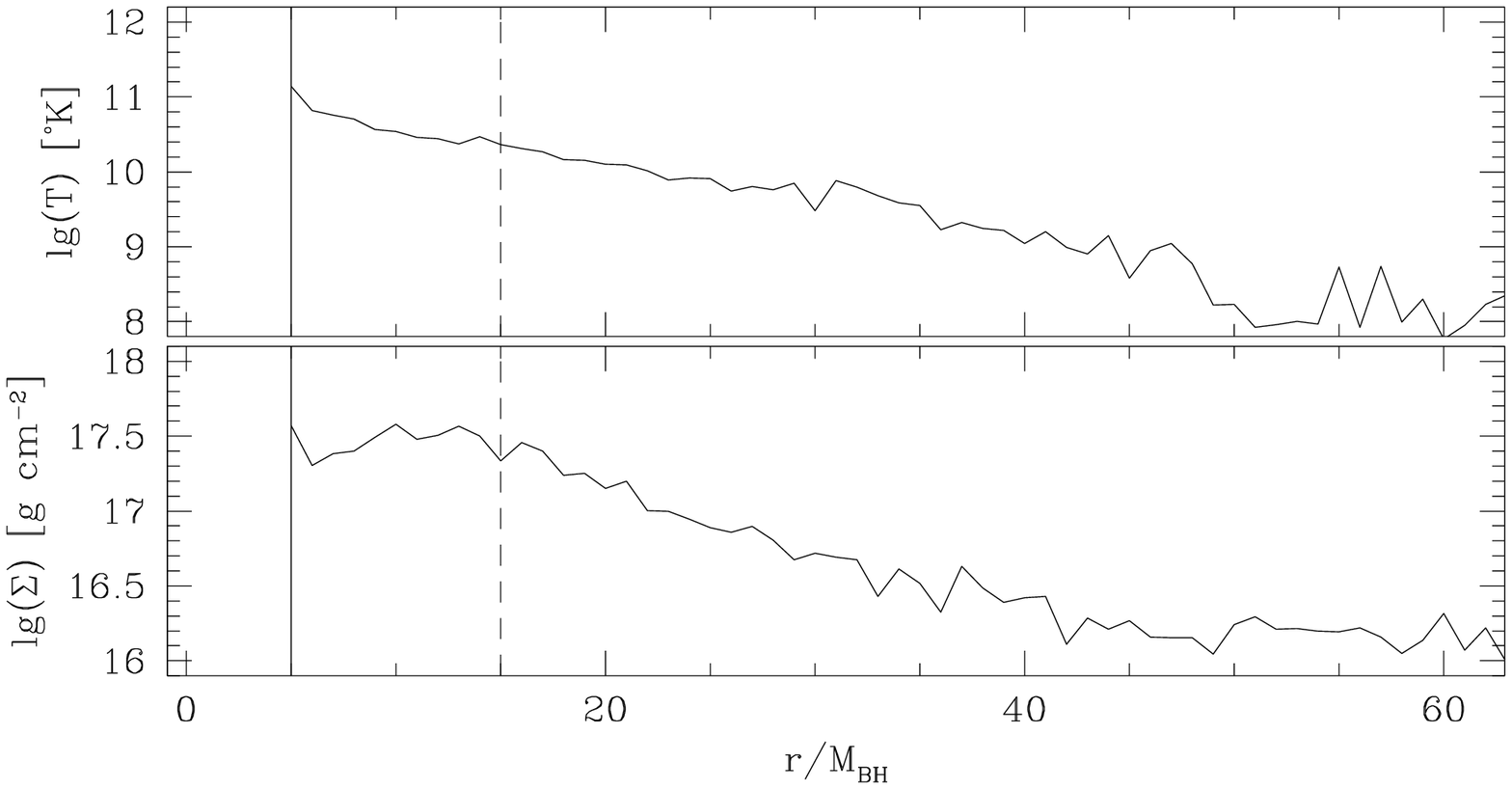}
\caption{Temperature $T$ and surface density $\Sigma$ as a function of
  cylindrical radius for the configuration shown in
  Fig.~\protect\ref{fig:xyz2}.  The solid vertical line denotes the
  ISCO, and the dashed line the transition radius between matter
  optically thick to neutrinos within and optically thin
  outside.\label{fig:rhoeps}}
\end{figure*}

The bound matter in the disk is relatively cold at first, as the
matter in the arm is initially ejected without
strong shock heating.  Over time,
this disk generates heat via shocks as matter falls back and 
wraps around the BH forming a torus
out to a radius of $r\sim 50M_{\rm BH}$ within $t=1000M_{BH}\sim
  .07~{\rm s}$; (see Fig.~\ref{fig:xyz2}).
The specific internal energy in the inner part of the torus
corresponds to a temperature $T\approx 3-10 MeV\approx 2-7\times
10^{10} K$; (see Fig.~\ref{fig:rhoeps}).  The surface density in this
region is $\Sigma\approx 2-3\times 10^{17} {\rm g~cm}^{-2}$.  Assuming
an opacity $\kappa=7\times 10^{-17}(T/10^{11} K)^2$ \citep{DPN}, we
conclude that the disk is optically thick out to $r\sim 15M$.  Using
the diffusion limit, the neutrino flux is given by $F_{\nu}\approx
(7N_{\nu}/3)(\sigma T^4/\kappa\Sigma)$ where $N_{\nu}=3$ is the number
of neutrino families, and $\sigma$ is the Stefan-Boltzmann constant.
The neutrino luminosity is $L_{\nu}\approx 2\pi r^2F_{\nu}\sim
10^{54}~{\rm erg~s}^{-1}$.  This value is roughly an order of
magnitude larger than that seen in the collapse of an HMNS
\citep{ShiUIUC}, and should produce at least a comparable annihilation
luminosity, $L_{\nu \bar{\nu}}\sim 10^{49-50}~{\rm erg~s}^{-1}$.

Qualitatively, the hot torus described here is similar both to the
initial data used to study hydrodynamic disk evolution in earlier GRB
models \citep{SRJ1,SRJ2,Aloy}, as well as to that formed by the
collapse of an HMNS \citep{ShiUIUC,Duez06}. However, we note that the
torus described here is physically larger.  Unlike
previous pseudo-Newtonian calculations of BHNS mergers
\citep{JERF,LK,RSW}, we find prompt disruption of the NS during
the plunge  (rather than a NS core which survives
the initial mass loss phase) and a lower mass disk, but at a
temperature similar to those calculations which included a detailed
microphysical treatment \citep{JERF}.

While we do not follow the long-term evolution of the accretion torus
and surrounding material, we can estimate the fallback time for the
bound component, assuming geodesic orbits.  Approximately $0.03M_{\rm
NS}$ should return back toward the BH on timescales equal to or longer
than a second, which could in principle produce lower-energy bursts at
later times.  It is conceivable that this fallback accretion might
explain the secondary X-ray flares observed in SGRBs many seconds
after the initial burst (see \citealp{Berger3} for a summary of the
observations), especially if self-gravity leads to the formation
of higher density clumps of material, but further simulations are
required to establish this identification.

Future observations should help to determine which progenitor
candidates are responsible for the observed short GRB population.
Gravitational wave measurements would provide important evidence if
detected in coincidence: A GRB resulting from hypermassive collapse
would occur noticeably delayed relative to the gravitational wave
signal from the inspiral and plunge phases of nearly equal-mass
objects, whereas one resulting from a BHNS binary would occur almost
immediately after the signal from a very unequal-mass merger.

\begin{acknowledgments}
It is a pleasure to thank Y.T.~Liu, M.~Shibata, and B.~Stephens for
useful conversations.  JAF is supported by an NSF Astronomy and
Astrophysics Postdoctoral Fellowship under award AST-0401533.  TWB
gratefully acknowledges support from the J.~S.~Guggenheim Memorial
Foundation.  This work was supported in part by NSF grants PHY-0205155
and PHY-0345151 and NASA Grant NNG04GK54G to the University of
Illinois, and NSF Grant PHY-0456917 to Bowdoin College.
\end{acknowledgments}

\end{document}